\documentstyle[psfig]{mn}

\title[The vertical composition of disc galaxies]{Tracing the vertical
composition of disc galaxies through colour gradients\thanks{Based on
observations obtained at the European Southern Observatory, La Silla,
Chile}}

\author[R.  de Grijs and R.F.  Peletier]{Richard de
Grijs$^1$\thanks{E-mail: grijs@virginia.edu} and Reynier F. 
Peletier$^{2,3}$ \\
$^1$ Astronomy Department, University of Virginia, PO Box 3818,
Charlottesville, VA 22903, USA \\
$^2$ Department of Physics, University of Durham, South Road, Durham
DH1 3LE \\
$^3$ School of Physics and Astronomy, University of Nottingham,
University Park, Nottingham NG7 2RD}

\date{Received date; accepted date}
\pubyear{1999}

\begin{document}
\maketitle

\begin{abstract}
Optical observations of a statistically complete sample of edge-on
disc-dominated galaxies are used to study the intrinsic vertical colour
gradients in the galactic discs, in order to constrain the effects of
population gradients, residual dust extinction and gradients in the
galaxies' metal abundance.  For the majority of our sample galaxies, the
colours and colour gradients in the range $1.0 h_z \le |z| \le 3.0 h_z$
most likely reflect the intrinsic galactic properties (where $h_z$ is
the vertical scale height).  \\
It appears that the intrinsic vertical colour gradients are either
non-existent, or small and relatively constant as a function of position
along the galaxies' major axes.  On average, the earlier-type galaxies
exhibit smaller vertical $(B-I)$ gradients than the later types; our
results are consistent with the absence of any vertical colour gradient
in the discs of the early-type sample galaxies.  In most galaxies
small-scale variations in the magnitude and even the direction of the
vertical gradient are observed: at larger galactocentric distances they
generally display redder colours with increasing {\it z} height, whereas
the opposite is often observed in and near the galactic centres.  \\
For a significant fraction of our sample galaxies another mechanism in 
addition to the effects of stellar population gradients is required to
explain the magnitude of the observed gradients. The non-zero colour
gradients in a significant fraction of our sample galaxies are likely 
(at least) partially due to residual dust extinction at these {\it z}
heights, as is also evidenced from the sometimes significant differences
between the vertical colour gradients measured on either side of the 
galactic planes. \\
We suggest that initial vertical metallicity gradients, if any, have
likely not been accentuated by accretion or merging events over the
lifetimes of our sample galaxies.  On the other hand, they may have
weakened any existing vertical metallicity gradients, although they also
may have left the existing correlations unchanged. 
\end{abstract}

\begin{keywords}
galaxies: abundances -- galaxies: fundamental parameters -- galaxies:
photometry -- galaxies: spiral -- galaxies: statistics
\end{keywords}

\section{Vertical colour and metallicity gradients}

The study of metal abundances in the disks of external spiral galaxies
has traditionally been performed using colours and colour gradients. 
Although, in theory, the use of line strengths allows to determine
stellar population parameters more accurately, the spatial and spectral
resolution required for this purpose are so large that thusfar such a
study has not been attempted. 

In contrast to the large number of studies of {\it radial} colour
gradients in moderately inclined and face-on spiral galaxies (e.g., de
Jong 1996, and references therein), the {\it vertical} colour behaviour
of highly inclined and edge-on galaxies has not received much attention. 
In highly inclined galaxies, the interpretation of intrinsic colours and
colour gradients is severely hampered by the presence of dust in the
galactic planes, which causes the dust lane to appear as a red feature
in vertical colour profiles (e.g., Hamabe et al.  1979, 1980; van der
Kruit \& Searle 1982a,b; Sasaki 1987; Wainscoat, Freeman \& Hyland 1989;
Aoki et al.  1991; Jansen et al.  1994; de Grijs, Peletier \& van der
Kruit 1997). 

In those galaxies, in which the dust lane is spatially resolved, the
presence of H{\sc ii} regions and a young, blue population of massive O
and B stars very close to the galactic plane, the {\sl ``young disc''},
can be inferred from the vertical colour profiles.  As was noted by,
e.g., van der Kruit \& Searle (1982b) for NGC 5907 and Wainscoat et al. 
(1989) for the southern edge-on galaxy IC 2531, the relative prominence
of the young disc in the bluer passbands, where reddening plays an
increasingly important role, means that it is very blue compared to the
dominant (``thin'') disc component.  The blue stars of which the light
is observed must therefore be located at sufficiently large radii, in
order to compensate for dust absorption.  The young stellar population
in our Galaxy also resides in a very thin young disc with a much smaller
scale height than that of the old-disc stars (e.g., Bahcall \& Soneira
1980). 

\subsection{Probing the composition of the Galaxy}

Because of our unique position within the Galaxy, studies of the
Galactic disc obviously provide the most detailed information about the
vertical composition of spiral galaxy discs.  On the other hand, our
location in the Galaxy also prevents us from probing its stellar
populations out to large distances along the Galactic plane, due to the
presence of interstellar extinction in the plane.  Studies of the
Galactic abundance distribution based on statistically complete samples
of tracer stars are therefore predominantly confined to lines of sight
(close to) perpendicular to the Galactic plane. 

For studies of the composition of the integrated stellar population in
the Galactic disc, the choice of a sample of representative tracer stars
is very important.  Ideally, one would like to select samples of
long-lived (i.e., of the order of the lifetime of the Galaxy), but
unevolved (main-sequence) stars that are easily observable.  Although
intrinsically bright tracer candidates are desirable in terms of
observing time, they suffer from a number of characteristics that make
them less suitable (see Gilmore, Wyse \& Jones 1995 for a discussion). 

K-type giants have been used extensively, but one must first identify K
giants from among the significantly larger number of K dwarfs with
similar apparent magnitudes and colours.  Therefore, F and early G-type
dwarfs are generally considered as the most desirable probes, because
they are the brightest of the unevolved stars that were formed during or
shortly after the formation epoch of the Galaxy.  Since their
atmospheres are well-mixed and they are ubiquitous throughout the
Galaxy, their chemical abundances provide us with a fossil record of the
integrated star formation and chemical enrichment history during the
early stages of the formation of the Galaxy.  Moreover, when using these
stars as tracers of the Galactic chemical enrichment history, one does
not need to apply model-dependent corrections for dead stars (cf.  van
den Berg 1962; Schmidt 1963; Tinsley 1980). 

\subsection{Metallicity gradients in the Galactic disc}
\label{metgrads.sect}

Although a large number of studies have focused on the presence and size
of a vertical metallicity gradient in the Galactic disc (e.g., Hartkopf
\& Yoss 1982; Grenon 1987; Yoshii, Ishida \& Stobie 1987; Yoss, Neese \&
Hartkopf 1987; Norris \& Green 1989; Majewski 1992; Reid \& Majewski
1993; J{\o}nch-S{\o}rensen \& Knude 1994; J{\o}nch-S{\o}rensen 1995;
Trefzger, Pel \& Gabi 1995; Robin et al.  1996; Buser, Rong \& Karaali
1998), the values for the actual decrease of mean metal abundance of the
old thin disc component with increasing distance from the Galactic plane
vary significantly.  This apparent disagreement is partly due to the
different sample selection criteria used to trace the abundance
gradient, as well as to differences in the definitions of the Galactic
disc components used by various authors. 

In particular, the observed metallicity gradients may be affected by the
admixture of the different disc components if the sample stars used to
trace these gradients are not easily separable for each of these
components.  Both Yoss et al.  (1987) and Majewski (1992) show that the
metallicity distributions of their sample F and G dwarfs at different
{\it z}-heights above the Galactic plane are non-Gaussian, implying a
mixture of stellar populations at all heights above the plane.  Yoss et
al.  (1987) argue that the thin young and the thin old disc are
essentially equally dominant for $z \le 250$ pc; the thin old disc is
most evident for $250 < z \le 500$ pc, whereas the thick disc starts to
show up for $500 < z \le 1000$ pc and dominates for $1000 < z \le 2200$
pc, above which the main contribution is that of the Galactic halo. 

Close examination of the available metal abundance data reveals that the
metallicity gradient is steeper within 800-1000 pc of the Galactic plane
than for $z > 1$ kpc (e.g., Hartkopf \& Yoss 1982; Yoss et al.  1987;
J{\o}nch-S{\o}rensen 1995; Trefzger et al.  1995), which corresponds to
a change in the relative dominance of the old thin and the thick disc
components. 

The mean abundance gradient for $z > 1$ kpc is of order $-0.18 \pm 0.05$
dex kpc$^{-1}$ (Hartkopf \& Yoss 1982; Yoshii et al.  1987; Yoss et al. 
1987; Norris \& Green 1989; Buser \& Rong 1995; J{\o}nch-S{\o}rensen
1995; Trefzger et al.  1995).  On the other hand, Majewski (1992) and
J{\o}nch-S{\o}rensen (1995) argue that their data are consistent with no
metallicity gradient at all in the thick-disc region, while the
non-Gaussian nature of the abundance distribution as a function of {\it
z}-height may cause the detection of an artificial gradient.  However,
Trefzger et al.  (1995) caution that the decreasing completeness of the
absolutely faintest stars in the tracer samples means that with
increasing distances one tends to lose preferentially the metal-poor
stars, which, in turn, causes the observed metallicity gradients to
flatten.  The good agreement between their and other studies leads them
to believe that this effect is insignificant, though. 

Most direct studies of the abundance gradient are restricted to stars in
the old thin disc, which is also the main disc component that we observe
in other galaxies.  Modern estimates of the mean metallicity gradient in
the $\sim 700 - 1000$ pc closest to the Galactic plane (Hartkopf \& Yoss
1982; Yoshii et al.  1987; Yoss et al.  1987; J{\o}nch-S{\o}rensen \&
Knude 1994; Buser \& Rong 1995; Trefzger et al.  1995; see also Robin et
al.  1996; Buser et al.  1998) are in the range $-0.6 \le \hbox{\rm
d[Fe/H]/d{\it z}} \le -0.3$ dex kpc$^{-1}$.  All of these results have
been obtained from {\it in situ} samples, along sight lines closely
perpendicular to the Galactic plane and centered on the position of the
Sun, thereby ignoring possible radial metallicity gradients. 
J{\o}nch-S{\o}rensen (1995) observed F and early G-type (main sequence)
stars in six selected directions of the Galaxy and tried to solve for
radial and vertical gradients simultaneously.  His best results for $z <
700$ pc are $-0.2 \pm 0.3$ dex kpc$^{-1}$ and $-0.01 \pm 0.03$ dex
kpc$^{-1}$ for the vertical and radial gradients, respectively. This may
indicate that the effects of such a small radial metallicity gradient on
the expected vertical abundance gradients are small or negligible, in
particular if the origin of either gradient is different.

\subsection{External galaxies: what do we know?}
\label{external.sect}

Colour distributions and colour gradients are sensitive to the metal
abundances and their gradients of the integrated stellar populations in
galaxies. 

Although {\it radial} colour gradients in moderately inclined and
face-on galaxies have been studied extensively, in general indicating
bluer colours with increasing galactocentric distance (e.g., de Jong
1996, and references therein), only for a few relatively large and
well-resolved edge-on galaxies {\it vertical} colour gradients have been
measured. 

In highly inclined galaxies, the interpretation of intrinsic colours and
colour gradients is severely hampered by the presence of dust in the
galactic planes.  However, from a comparison with published colours of
moderately inclined Sc galaxies, Kuchinski \& Terndrup (1996) have shown
that for these late-type galaxies there is little or no reddening away
from the dust lane.  Since statistical studies have shown that the dust
content of Sc galaxies is large compared to other disc-dominated galaxy
types (e.g., de Grijs et al.  1997), we may assume that the effects of
reddening on the intrinsic galaxy colours away from the dust lane are
largest for these galaxy types.  Thus, colours and colour gradients
measured at those distances from the galactic planes where the influence
of the dust lane is negligible likely reflect the {\it intrinsic}
galactic properties.  Detailed studies of the intrinsic colours of
galactic discs perpendicular to their planes (e.g., Hamabe et al.  1979;
Hegyi \& Gerber 1979; van der Kruit \& Searle 1981; Jensen \& Thuan
1982) are consistent with a small or no vertical colour gradient outside
the dust lane region (see also de Grijs et al.  1997). 

Although colour gradients along the minor axis may be due to some
intrinsic bulge property\footnote{However, Peletier \& Balcells (1996,
1997) showed, for their sample of 30 field spiral and lenticular
galaxies, that the bulge colours on the minor axis and the inner disc
colours taken in wedge apertures at 15$^\circ$ from the major axis at 2
{\it K}-band scale lengths are very similar.}, van der Kruit \& Searle
(1981) observed that, at various galactocentric distances, the vertical
colours of NGC 891 are getting systematically bluer with greater height
above the plane.  On the other hand, Jensen \& Thuan (1982) did not find
any evidence for a similar vertical colour gradient in NGC 4565 in the
region where the old thin disc dominates.  However, as soon as the light
of the thick disc starts to dominate a small perpendicular colour
gradient is present in their data, in the sense that the disc colours
become redder with increasing distance from the galactic plane.  A
similar result has recently been obtained for NGC 5907 (Lequeux et al. 
1996, 1998; Rudy et al.  1997), which was interpreted as an extended
stellar halo redder than the galactic disc or a very thick disc
component. 

The conversion of broad-band colour gradients to abundance and
population gradients in external galaxies is controversial,
unfortunately.  For the detailed analysis of the luminosity and colour
profiles of edge-on galaxies one needs to adopt {\it a priori}
assumptions concerning the evolutionary stellar population synthesis,
the initial mass function, the metallicity and the star formation
history, as well as about the dust geometry and its characteristics. 
Due to the relative insensitivity of broad-band colours to these
characteristics, in particular because of the age/metallicity degeneracy
in the colours of an integrated stellar population (Worthey 1994),
spectral line studies seem to be a more effective tool to disentangle
metallicity and age effects, as well as population gradients. 

However, spectral line strength indices are relatively hard to measure,
and are also degenerate to age and metallicity, although to a lesser
extent than broad-band colours.  Since {\it local} colours correlate
strongly with each other (de Jong 1996, Peletier \& Balcells 1996, 1997)
they can be used as indicators of the gross properties of galaxies, in
the absence of dust, where the various wavelength ranges can be used as
diagnostics for different overall galaxy properties.  In this respect,
de Jong's (1996) statistical study is one of the first large surveys of
spiral galaxy properties based on multiple passband optical and
near-infrared observations. 

Fisher, Franx \& Illingworth (1996) published one of the very few
studies dealing with abundance gradients perpendicular to the galactic
planes in highly-inclined galaxies other than our own.  Based on Mg$_2$
spectral line observations of 20 S0 galaxies, they conclude that the
minor axis behaviour of the 9 galaxies in their edge-on subsample is
noticeably different from that found along the major axis.  Whereas the
major axis Mg$_2$ profiles decrease with radius and flatten as the bulge
light contribution decreases and the disc starts to dominate, the minor
axis gradients display a uniformly decreasing Mg$_2$ strength with
distance from the galactic plane, reaching lower values than those of
the major axis metallicities. 

\section{Approach}

\subsection{Sample selection, observations, and reduction techniques}

The selection of our statistically complete sample of edge-on
disc-dominated galaxies, and the (optical) observations and data
reduction techniques on which the results presented in this paper are
based were described in detail in de Grijs (1997, 1998) and de Grijs \&
Peletier (1999).  Here, we summarize the selection criteria that were
applied to the galaxies in the Surface Photometry Catalogue of the
ESO-Uppsala Galaxies (ESO-LV; Lauberts \& Valentijn 1989):

\begin{enumerate}
\item Inclination selection: $i \ge 87^\circ$;
\item Diameter selection: $D_{25}^B \ge 2.'2$ ($D_{25}^B$: blue angular
size at a surface brightness $\Sigma_B = 25$ mag arcsec$^{-2}$);
\item Galaxy type selection:  S0 -- Sd, and
\item Morphology selection: non-interacting and undisturbed galaxies.
\end{enumerate}

The ESO-LV is statistically complete for diameter-limited samples with
$D_{25}^B \ge 1.'0$ (de Grijs 1997), based on $V/V_{\rm max}$
completeness tests (e.g., Davies 1990; de Jong \& van der Kruit 1994). 
Our observed subsample is characterized by $V/V_{\rm max} = 0.502 \pm
0.253$ (for $D_{25}^B \ge 2.'2$), which implies statistical completeness
for the criteria outlined above. 

In de Grijs (1997, 1998), we compared our photometry to that of de
Vaucouleurs et al.  (1991; RC3), Mathewson, Ford \& Buchhorn (1992) and
Mathewson \& Ford (1996), and Barteldrees \& Dettmar (1994).  From the
detailed comparison of our photometry to theirs, using both luminosity
profiles and total magnitudes, it was shown that we can accurately
reproduce their results (e.g., for the comparison of total magnitudes we
find $\langle m_{I,{\rm our}} - m_{I,{\rm Mathewson}}\rangle = -0.07 \pm
0.13$, and $\langle m_{B,{\rm our}} - m_{B,{\rm RC3}}\rangle = -0.09 \pm
0.45$). 

\subsection{The determination of vertical colour gradients}
\label{determination.sect}

The determination of colours and colour gradients in our sample galaxies
requires a careful matching of the individual observations taken through
different filters, or with different telescope/instrument combinations
on different nights. 

First, if our multi-passband observations of a single galaxy were
obtained with different telescope/instrument combinations, we rebinned
the images to a common reference frame (under conservation of the
observed flux), such that the highest-quality observations (generally
those taken with the Danish 1.54m telescope), or the largest number of
observations that were taken with the same telescope/instrument
combination, remained unaffected. 

The rebinning and the alignment of the individual images were performed
using the standard IRAF\footnote{The Image Reduction and Analysis
Facility (IRAF) is distributed by the National Optical Astronomy
Observatories, which is operated by the Association of Universities for
Research in Astronomy, Inc., under cooperative agreement with the
National Science Foundation.} tasks {\sc magnify, rotate}, and {\sc
imalign}.  We determined the rotation angles and alignment offsets using
common (foreground) stars in the individual frames.  Only in those rare
cases where foreground stars were completely lacking the galactic
centres were used for the alignment. 

In order not to be hindered by contamination of the galaxy luminosity by
foreground star light, we used the {\it I}-band images to mask out those
areas contaminated by foreground stars.  We used the mask thus obtained
to correct for foreground star contamination in all passbands. 

Secondly, we extracted vertical luminosity profiles at a number of
positions along the major axes of the sample galaxies.  A
semi-logarithmic binning algorithm was applied to the galaxies, both
radially and vertically, to retain an approximately constant overall
signal-to-noise (S/N) ratio in the resulting vertical profiles.  This
binning algorithm was applied to all individual passbands of a given
galaxy with exactly identical parameters, to facilitate the subsequent
construction of the vertical colour profiles. 

We rejected those profiles with poor S/N ratios (generally the outermost
profiles) and those that were clearly affected by artifacts in the data
or foreground stars.  For all galaxies we were able to sample the
vertical luminosity distribution at various positions along the major
axis outside the region where the bulge contribution dominates.  For
galaxies of types $T \ge -1.0$ we could determine the luminosity
profiles for at least 4 of these independent positions in each passband. 

The photometric calibration of the luminosity profiles was done using
the calibration parameters obtained in de Grijs (1997, 1998). 

Finally, we combined the calibrated vertical luminosity profiles
obtained in the individual passbands into our final, calibrated colour
profiles, in $(B-V)$, $(V-I)$, and $(B-I)$.  Due to the low S/N ratios
and the snapshot nature (and the resulting shallow profiles) of our
near-infrared $K'$-band observations (de Grijs 1997; de Grijs \&
Peletier 1999), we will not discuss any optical/near-infrared colour
gradients. 

The positions of the galactic centres and the galactic planes were
determined by folding the profiles and under the assumption of
symmetrical light distributions with respect to the planes. 

Colour gradients in all our vertical colour profiles were obtained by
fitting straight lines to the data points within user-specified vertical
ranges (in units of the galaxy's scale height), using a reduced $\chi^2$
least-squares minimization routine.  In all our sample galaxies, we
determined the presence and magnitude of the possible colour gradients
at each sampled position by examining the colour profiles on each side
of the galactic planes separately.  In the following section, we will
discuss the effects of choosing different vertical fitting ranges, and
present the results for our entire sample of edge-on disc galaxies. 

\section{Statistical trends}

\subsection{The extent of the dust lane: choosing the vertical fitting
range} 

The correct choice of the vertical fitting range to study the presence
of possible colour gradients is important for two reasons:

\begin{enumerate}

\item At small distances from the galactic planes, the observed colours
are affected by the in-plane interstellar extinction in the galaxies,
causing the dust lane to appear as a red feature in vertical colour
profiles (Sect.  \ref{external.sect}), superposed on the intrinsically
bluer disc population.  Thus, choosing a lower fitting limit too close
to the galactic planes will artificially increase the measured slope and
result in an erroneous interpretation of the colour gradient. 

\item Fortunately, however, as already mentioned in Sect. 
\ref{external.sect}, Kuchinski \& Terndrup (1996) have shown that there
is little or no reddening away from the dust lane in their sample of Sc
galaxies, which are typically the dustiest galaxy types.  Therefore,
colours and colour gradients measured at those distances from the
galactic planes where the influence of the dust lane is negligible
likely reflect the {\it intrinsic} galactic properties.  However, at
large {\it z} distances, the S/N ratios of the individual luminosity
profiles, from which the colour profiles were constructed, decrease
significantly, causing the S/N ratios of the colour profiles to become
even poorer. 

\end{enumerate}

Obviously, any choice of the vertical fitting range that will give us a
good representation of the vertical colour gradient at any one position
will therefore include only those data that are mostly unaffected by the
dust lane, but of which the S/N ratio is still sufficient (we only used
luminosity profiles with S/N $> 5$ in the fitting ranges used to
construct the colour profiles). 

In de Grijs et al.  (1997) we showed that the vertical extent of the
dust lane in our sample galaxies is clearly revealed by the $(I-K)$
minor axis colour profiles presented in that paper.  These colour
profiles likely indicate the maximum extent of the dust lanes, since its
maximum effect is expected in the galactic centres (cf.  the colour maps
in de Grijs 1997 and de Grijs et al.  1997).  From the $(I-K)$ minor
axis colour profiles in de Grijs et al.  (1997), we estimate that -- for
the majority of our sample galaxies -- we can use a lower limit $z_{\rm
lower} \simeq 1.0 - 1.5 h_z$ (where $h_z$ is the scale height of the
dominant old thin disc component), and an upper limit $z_{\rm upper}
\simeq 2.5 - 3.0 h_z$.  The determination of the scale heights of our
sample galaxies was discussed in detail by de Grijs et al.  (1997) and
de Grijs (1998). Briefly summarized, we fitted the generalized vertical
density laws proposed by van der Kruit (1988) to the vertical luminosity
profiles of the sample galaxies for which we had $K'$-band observations
available; for the other galaxies we used an iterative fitting routine
to determine the slope of their vertical {\it I}-band luminosity
distributions between 1.5 and 4.0 $h_z$. In both cases, we decided to
use our longest-wavelength observations to best approximate the vertical
luminosity (and presumably mass) distributions (de Grijs et al. 1997, de
Grijs 1998).

By choosing various fitting ranges within these boundaries and by
treating both sides of the galactic planes separately we can exclude
profiles that are affected by any remaining extinction effects due to
the dust lane and by any effects due to poor S/N ratios at high {\it z}
heights (or large radial distances).  In addition, this method will help
us to filter out the effects of foreground stars, stellar warps and
patchy spiral arms. 

\subsection{Vertical colour gradients}
\label{vertcols.sect}

To obtain estimates of the colour gradients that are least affected by
artifacts due to extinction or poor S/N ratios, we applied our colour
gradient fitting routine to the sample galaxies using three vertical
fitting ranges: $|1.0 - 2.5| h_z, |1.0 - 3.0| h_z$ and $|1.5 - 3.0|
h_z$, for all three colours discussed in Sect. \ref{determination.sect}. 

Close examination of the results obtained from these three fitting
ranges led us to conclude that the total acceptable range, $1.0 h_z \le
|z| \le 3.0 h_z$, produced -- in general -- the most representative
vertical colour gradients as a function of projected galactocentric
distance: any small-scale variations due to excess out-of-plane
extinction, foreground star light or artifacts due to low S/N ratios in
the outer regions are smoothed out by this relatively large vertical
range, which we will use in the subsequent colour gradient analysis. 
From the point-to-point variation between adjacent data points, it is
clear that the results obtained from both the inner range, $1.0 h_z \le
|z| \le 2.5 h_z$, and the outer range, $1.5 h_z \le |z| \le 3.0 h_z$
show greater scatter than those of the total range. 

Next, we examined the individual colour profiles and their corresponding
colour gradients for each of our sample galaxies visually, combined with
the original galaxy images, in order to assess the quality of the fits
and the usefulness of the profiles for our purposes.  The adoption of
this quality evaluation procedure enabled us to reject those profiles
that were clearly affected by dust (either because of a very prominent
dust lane or due to residual dust features), foreground stars, inclined
spiral arms, or very low S/N ratios. 

As a result, the colour profiles for eight of our sample galaxies (ESO
321-G10, ESO 340-G08, ESO 377-G07, ESO 435-G50, ESO 444-G21, ESO
506-G02, ESO 555-G36, and ESO 575-G61) proved to be too noisy, or
dominated by the colours of the sky background at $1.0 h_z \le z \le 3.0
h_z$, due to the shallow nature of the individual luminosity profiles
used to construct the colour profiles. In addition, for a number of
galaxies, we were forced to adjust the vertical fitting range because of
various reasons (e.g., significant reddening due to dust extending into
the adopted fitting range, a limited number of data points to produce
reliable fits, etc.).

\begin{figure*}
\vspace*{-1cm}
\psfig{figure=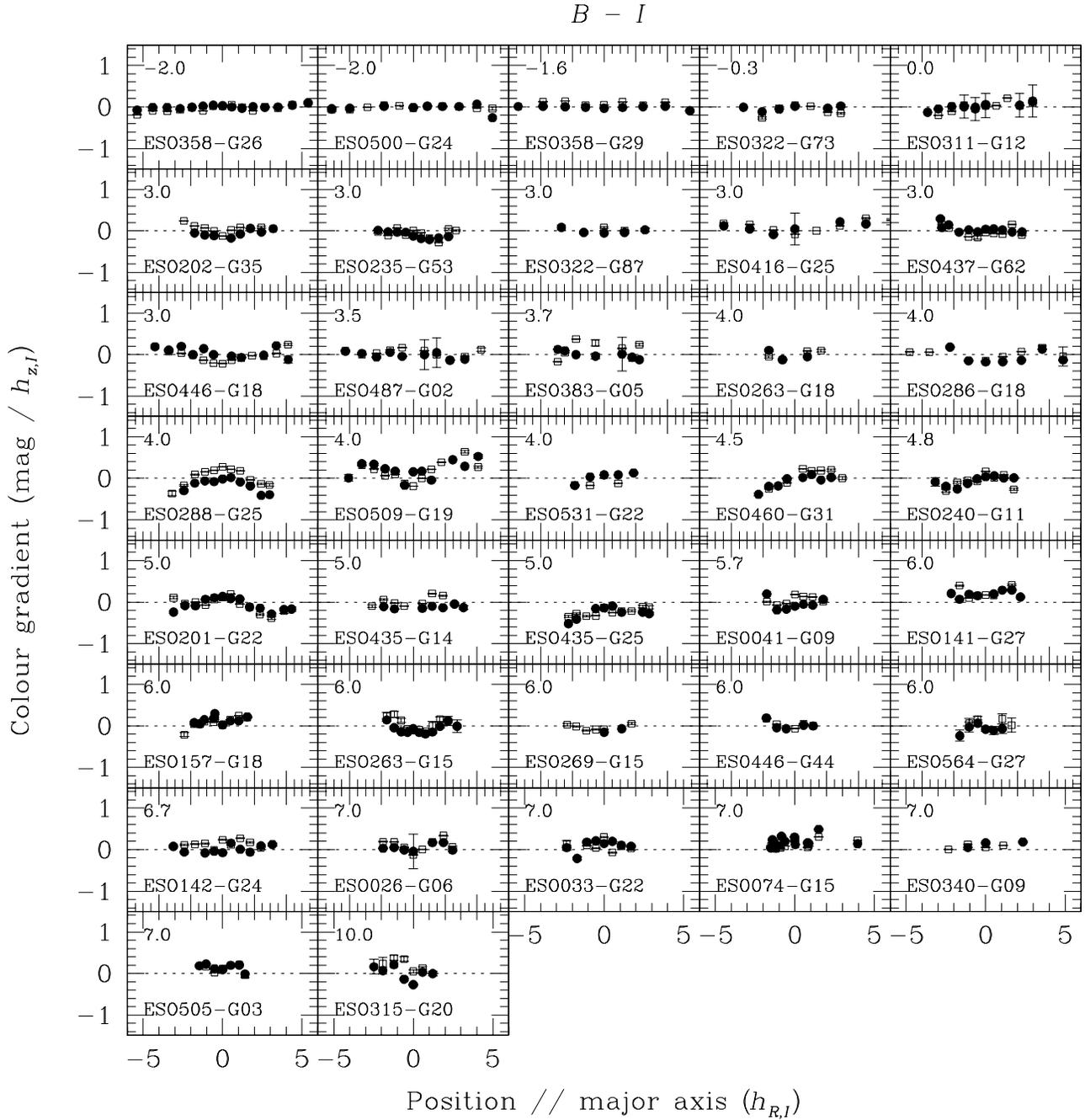,width=19cm}
\caption[]{\label{totgrads.fig}Vertical ($B-I$) colour gradients as a
function of projected galactocentric distance for all of our sample
galaxies, and sorted by (revised) Hubble type, {\it T}, indicated in the
upper left-hand corner of each panel.  We have plotted the results
obtained from the side of the plane most affected by extinction by open
squares and the other side by filled circles.  The fits were done in the
range $1.0 h_z \le |z| \le 3.0 h_z$, except for ESO 141-G27 ($|1.0 -
3.5| h_z$), ESO 201-G22 ($|1.0 - 3.5| h_z$), ESO 240-G11 ($|1.5 - 4.0|
h_z$), ESO 263-G15 ($|1.5 - 3.5| h_z$), ESO 311-G12 ($|1.0 - 4.0| h_z$),
ESO 322-G87 ($|1.5 - 3.0| h_z$), ESO 340-G09 ($|0.5 - 1.5| h_z$), ESO
435-G25 ($|1.5 - 3.5| h_z$), and ESO 446-G18 ($|1.5 - 3.5| h_z$).}
\end{figure*}

In Fig.  \ref{totgrads.fig} we show the resulting colour gradients as a
function of projected galactocentric distance for our largest colour
baseline, $(B-I)$, for all of our sample galaxies, and sorted by
(revised) Hubble type, {\it T} (indicated in the upper left-hand corner
of each panel).  We have plotted the results obtained from either side
of the plane by open squares and filled circles, respectively.  The
latter represent the side of the galactic plane least affected by
extinction, if appropriate.  However, one should note that in some cases
the dust lane corrugates, in which case the least (most) dusty side
refers to the side where the majority of the profiles are least (most)
affected by extinction.  The fits were done in the range $1.0 h_z \le
|z| \le 3.0 h_z$; for exceptions see the caption of Fig. 
\ref{totgrads.fig}. 

Although one can immediately see that the detected vertical colour
gradients are small and relatively constant as a function of position
along the galaxies' major axes, in most galaxies small-scale variations
in the magnitude, and even the sense of the colour gradients are
observed.  The differences are in the sense that the outer regions
generally display redder colours with increasing {\it z}-distance,
whereas the opposite behaviour is often observed in and near the
galactic centres.  In addition, it is also clear that although the
results obtained from either side of the galactic planes agree
reasonably well, small differences between both sides are appreciated in
almost all of our sample galaxies.  These are unlikely to be caused by
incorrect background subtractions, since we are comparing {\it relative}
colours within a given galaxy.  In all cases, the background emission in
the field of view of our programme galaxies could be well represented by
a two-dimensional plane, determined by the flux in regions sufficiently
far away from the galaxies themselves in order not to be affected by
residual galactic light.  For the majority of our observations, these
planes were closely approximated by constant flux values across the CCD
field.  The remaining uncertainties in the background contribution are
due to poisson noise statistics (see de Grijs [1997] for a detailed
overview of the image reduction techniques used). 

We will discuss the observed colour behaviour for those sample galaxies
for which we have high-quality colour profiles, and those that show
significant point-to-point variations, either on one side or compared to
the other side of the plane, or significant non-zero gradients.

For the detailed (projected) two-dimensional behaviour of colour as a
function of position within our sample galaxies, we refer the reader to
the $(I-K)$ colour maps for 24 of the galaxies in de Grijs et al. 
(1997) and the $(B-I)$ colour maps for all galaxies in de Grijs (1997,
Chapter 9). 

In Fig.  \ref{colprofs.fig} a number of the colour profiles, extracted
at various positions along the galaxies' major axes, are shown for a few
of the most instructive sample galaxies.  The calibrated colours apply
to the colour profiles at the galactic centres; offsets in colour, in
increments of $\Delta (B-I) = \pm 1.0$ mag, have been applied to the
other profiles for reasons of clarity and display purposes. 

\begin{itemize}

\item {\sl ESO 033-G22} is a very regular, late-type galaxy with a clear
dust lane that obscures the galactic centre on one side of the plane. 
Since it resides in a field with a large number of foreground stars, the
erratic behaviour of the vertical colour profiles is likely caused by
the superposition of these on the galactic light profiles.  (T = 7.0)

\item The vertical colour profiles on either side of the plane of {\sl
ESO 041-G09} show a significant difference, due to a broad dust lane on
one side of the galaxy; a number of the colour profiles are
significantly affected by foreground stars.  (T = 5.7)

\item {\sl ESO 074-G15} is a very late-type galaxy with irregular dust
patches scattered across its disc.  It does not display a clear galactic
centre, and its disc shows significant, irregular thickening on one side
of the galactic centre.  Its large projected size lends it very well for
a detailed study of its colour behaviour.  (Fig.  \ref{colprofs.fig}a; T
= 7.0)

\item Due to the relative absence of foreground stars around {\sl ESO
141-G27}, except for two bright stars that are viewed on top of the
galaxy, and because of its almost perfect edge-on orientation, the
vertical colour profiles on either side of the plane show a symmetrical
behaviour.  The colours become redder with increasing distance from the
plane; based on its appearance, we cannot rule out that this is due to a
small out-of-plane projection of the dust in the spiral arms. (T = 6.0)

\item The colour profiles on one side of the plane of {\sl ESO 142-G24}
are clearly affected by its dust lane; the side that is relatively
unaffected by extinction effects does not show any significant or
systematic vertical colour gradient. (T = 6.7)

\item Despite its regular appearance, {\sl ESO 157-G18} is clearly dusty
and shows irregularly located regions of high extinction throughout its
disc, which may be the cause of the small vertical colour gradient,
indicating redder colours with increasing {\it z} height.  (Fig. 
\ref{colprofs.fig}b; T = 6.0)

\item Although {\sl ESO 201-G22} shows some evidence for a patchy dust
distribution on its dusty side, its colour gradients on the side that is
relatively unaffected by extinction also show redder colours with height
above the plane. (T = 5.0)

\item {\sl ESO 202-G35} is one of our least inclined sample galaxies; it
clearly shows its projected spiral arms and the dust lanes associated
with these.  Any measured vertical colour gradients are therefore
compromised by extinction, rendering a direct interpretation of its
colours troublesome. (T = 3.0)

\item The highly warped galaxy {\sl ESO 235-G53} shows a clean-cut dust
lane that divides the galactic bulge neatly in two (almost) equal parts. 
Apart from this dust lane, the colour profiles are hardly affected by
residual dust extinction, but the warping of its disc requires a careful
modeling of its plane.  Its ``dust-free'' colours indicate a blueing of
the disc with increasing {\it z} height.  (Fig.  \ref{colprofs.fig}c; T
= 3.0)

\item {\sl ESO 240-G11} is a beautiful example of a disc-dominated
system that is unfortunately clearly affected by patches of dust across
its disc.  The side that is relatively little affected by the ubiquitous
extinction indicates a small (or even the absence of a) vertical colour
gradient. (T = 4.8)

\item {\sl ESO 263-G15} shows an irregular dust lane that cuts through
the middle of the galactic centre, where we observe bluer colours with
increasing distance from the plane, which are likely little affected by
extinction.  At larger galactocentric distances the colour profiles
change to indicate redder colours with height, which is almost certainly
due to the patchy extinction observed at these radii, as well as the
presence of young stars, and thus active star formation, close to the
galactic plane.  (T = 6.0)

\item Much of the scatter and noise in the vertical colour profiles and
the derived gradients in {\sl ESO 286-G18} is likely due to the low S/N
in our {\it B}-band image.  Colour profiles can be obtained with
acceptable accuracy only at radii relatively close to the galactic
center; they are consistent with the absence of any vertical colour
gradient. (T = 4.0)

\item Almost all of the outer-disc colour profiles of {\sl ESO 288-G25}
that are mostly unaffected by dust indicate redder colours with
increasing height above the plane; in these profiles, the effects of
dust are virtually non-existent in the {\it z} range used for our
analysis. (T = 4.0)

\item {\sl ESO 311-G12} is a very smooth-looking early-type spiral that
does not show any significant extinction effects. Although it resides in
a region that is relatively heavily populated by foreground stars, its
vertical colour profiles are hardly affected by those; they are
consistent with no colour gradient at all over the {\it z} range from
1.0 to 3.0 $h_z$. (Fig.  \ref{colprofs.fig}d; T = 0.0) 

\item Our low S/N {\it B}-band observation of {\sl ESO 315-G20}, which
is further compromised by a very bright foreground star located just
outside the main bulge region, is the likely cause for the noisy
behaviour of the magnitude of the vertical colour gradient with
galactocentric distance.  At those radii where the S/N is sufficient to
accurately examine the vertical gradient, our results indicate that any
such gradient is probably non-existent. (T = 10.0)

\item {\sl ESO 322-G73} is an early-type galaxy with a small
disc-component, which shows a clearly disturbed morphology, that may be
due to an interaction with a nearby galaxy in the same group. Apart from
the red feature associated with the in-plane dust lane, the dust-free
regions in its vertical colour profiles are essentially featureless and
consistent with a zero gradient. (T = $-0.3$)

\item {\sl ESO 358-G26} and {\sl ESO 358-G29} are among the earliest
galaxy types in our sample, and do not show any evidence for a sizable
amount of extinction.  The vertical colour profiles are featureless and
entirely consistent with no gradient at all.  ({\sl ESO 358-G26:} Fig. 
\ref{colprofs.fig}e, T = $-2.0$; {\sl ESO 358-G29:} T = $-1.6$)

\item Our observations of {\sl ESO 383-G05} show that although the inner
part of the galaxy is well-behaved, showing a bright bulge divided by a
prominent dust lane, the outer regions are highly irregular.  It appears
that an out-of-plane spiral arm obscures part of the main stellar disc
component, which is undoubtedly the cause for the erratic behaviour of
the open symbols in Fig.  \ref{totgrads.fig}.  The other side of the
plane appears to be very regular and hardly affected by extinction or
warping; the corresponding vertical colour gradients do not show any
significant deviation from zero. (T = 3.7)

\item {\sl ESO 416-G25} is characterized by a very thin disc and a clear
dust lane that divides the bulge in two equal parts, leaving patches of
dust in the disc at larger galactocentric distances.  Its vertical
colour gradients seem to indicate redder colours with increasing height
above the plane, similarly on either side of the plane. (T = 3.0)

\item {\sl ESO 435-G25} consistently shows bluer colours with increasing
{\it z} height, although we cannot determine whether we are observing
the intrinsic galaxy colours or the effects of residual dust extinction
at progressively higher positions in the disc.  The galaxy shows a very
prominent dust lane, extending from the galactic centre to the edge of
the disc, as well as into the higher stellar layers, as is evidenced by
dust features that can be traced up to a few scale heights.  (Fig. 
\ref{colprofs.fig}f; T = 5.0)

\item Outside the region where its dust lane creates a red feature in
its vertical colour profiles, {\sl ESO 437-G62} shows essentially
unchanging colours with increasing height above the plane.  The galaxy
image looks very smooth and does not show any evidence for residual
extinction effects.  (Fig.  \ref{colprofs.fig}g; T = 3.0)

\item The effects of dust in {\sl ESO 446-G18} are obvious if one
examines the appropriate panel in Fig.  \ref{totgrads.fig}: the side
that is most affected by extinction shows a clear change in the sense of
the vertical colour gradient, whereas the colour gradient on the other
side remains relatively constant and close to zero at those radii where
the effects of residual extinction are negligible.  (Fig. 
\ref{colprofs.fig}h; T = 3.0)

\item The colours of {\sl ESO 460-G31} are significantly affected by
extinction over the entire disc-dominated radial range, which make a
direct interpretation of the vertical colour gradients unreliable. (T =
4.5) 

\item {\sl ESO 487-G02} shows a behaviour of its colour profiles that is
remarkably similar to that shown by {\sl ESO 437-G62}; its vertical
colour profiles are consistent with a zero gradient outside the region
dominated by the strong dust lane in the plane. (T = 3.5)

\item As one of our earliest-type sample galaxies, {\sl ESO 500-G24}
does not show any evidence for a significant amount of extinction, nor
does it show any noticeable vertical colour gradient at any one position
along its major axis. (T = $-2.0$)

\item {\sl ESO 505-G03} is clearly a disturbed galaxy with a patchy dust
content and a thin disc.  However, based on its appearance, the
reddening of its colours with increasing {\it z} height might well be
real and not due to extinction effects. (T = 7.0)

\item Finally, {\sl ESO 509-G19} shows redder colours with increasing
distance from the plane.  Visual examination of the individual
observations lead to the conclusion that this can be attributed to
residual dust effects at these {\it z} heights. (T = 4.0)

\end{itemize}

\begin{figure*} 
\vspace*{-0.5cm} 
\psfig{figure=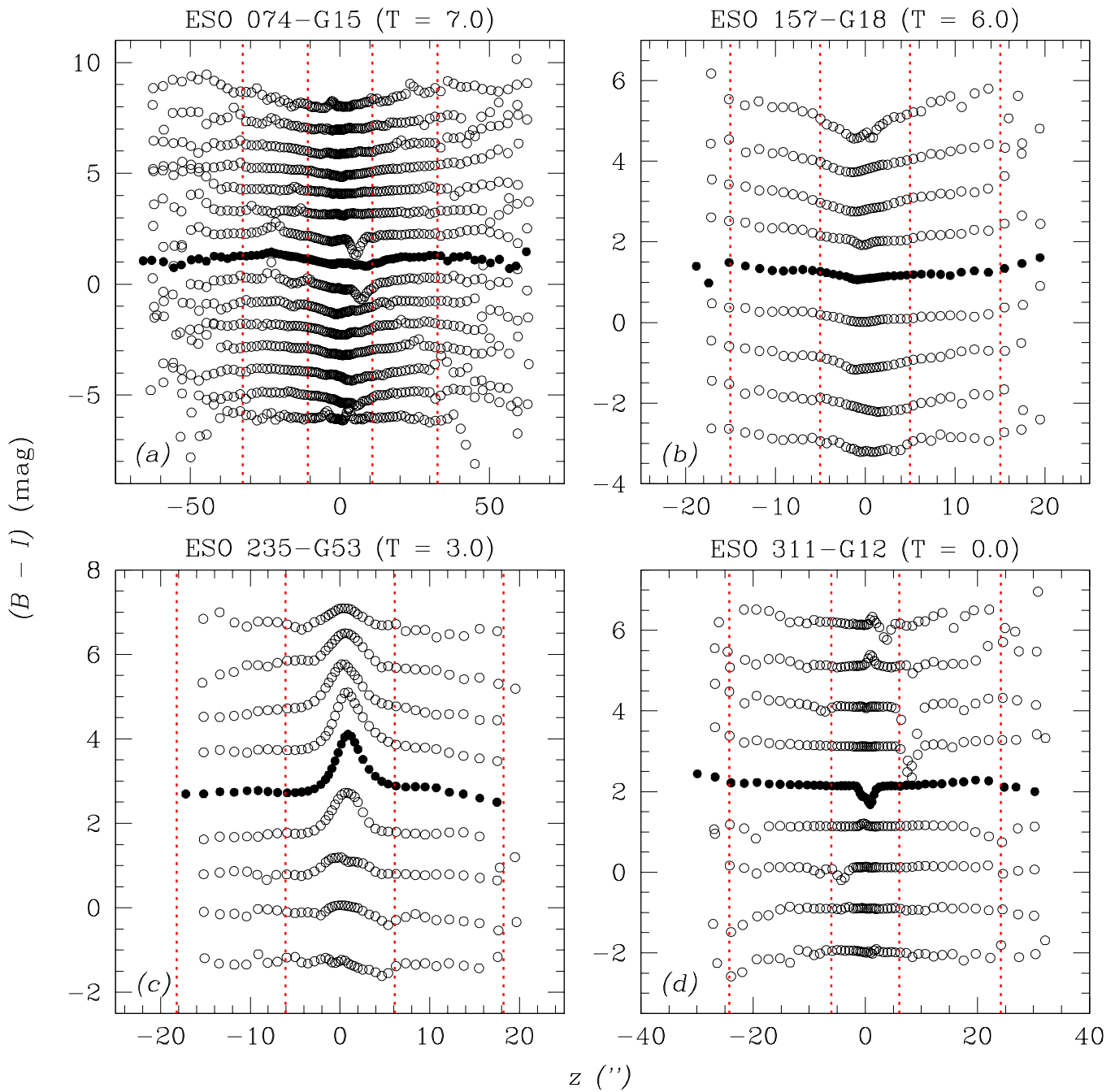,width=19cm}
\caption[]{\label{colprofs.fig}Vertical $(B-I)$ colour profiles of a
number of the most interesting and instructive sample galaxies.  The
calibrated colours refer to the central profiles (filled symbols); the
other profiles have been offset by $\Delta (B-I) = \pm 1.0$ mag for
display purposes and reasons of clarity.  The vertical dotted lines
indicate the fitting range used for each galaxy. {\it (a)} ESO 074-G15;
{\it (b)} ESO 157-G18; {\it (c)} ESO 235-G53; {\it (d)} ESO 311-G12.}
\end{figure*}

\addtocounter{figure}{-1}
\begin{figure*} 
\vspace*{-0.5cm} 
\psfig{figure=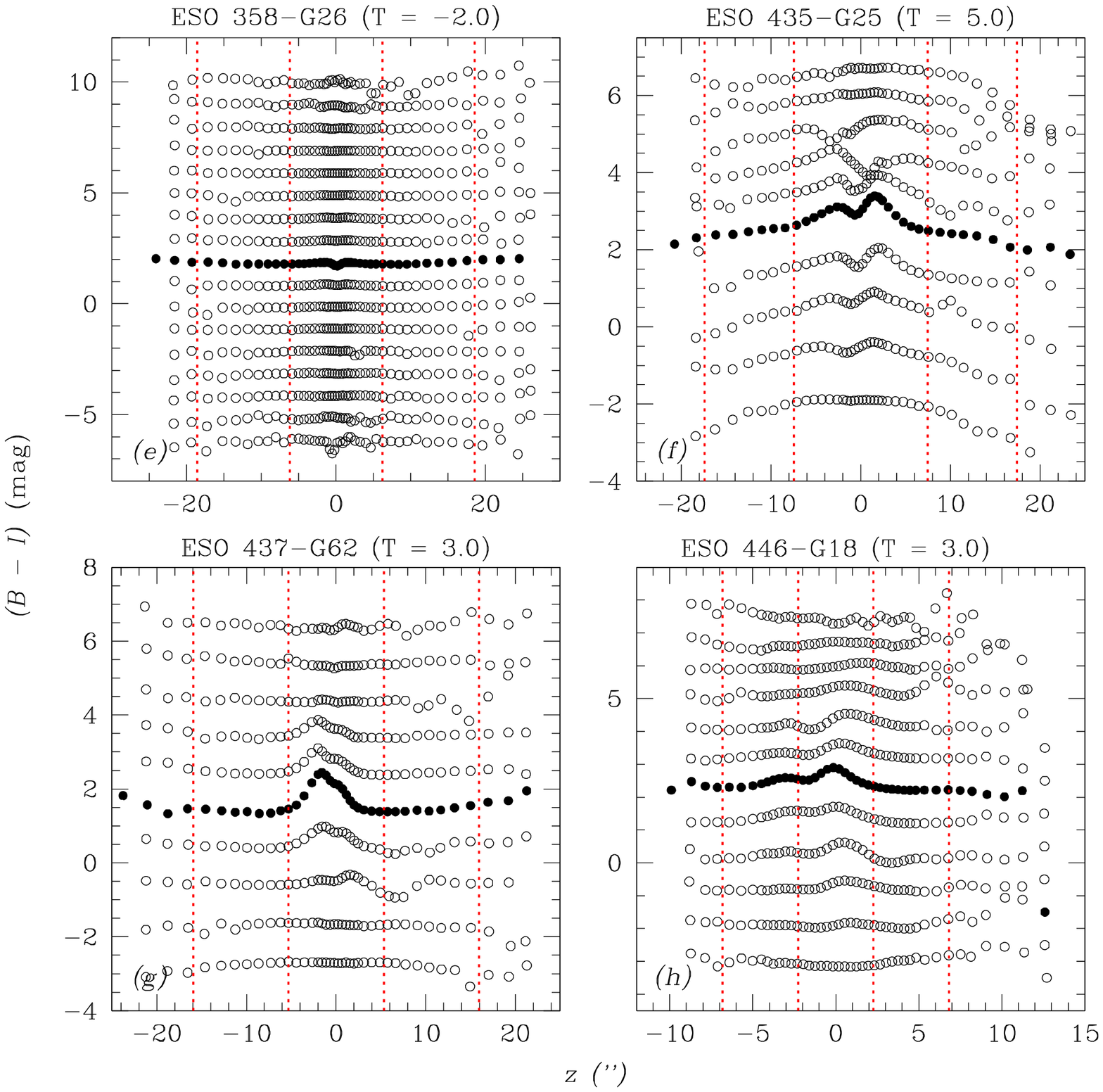,width=19cm}
\caption[]{{\sl (continued)} {\it (e)} ESO 358-G26; {\it (f)} ESO
435-G25; {\it (g)} ESO 437-G62; {\it (h)} ESO 446-G18.}
\end{figure*}

In summary, from the detailed examination of the vertical colour
profiles (Figs.  \ref{totgrads.fig} and \ref{colprofs.fig}) it appears
that the vertical profiles, at least those extracted on the side of the
galactic plane that is least affected by extinction, are consistent with
no gradient at all.  Although the individual panels of Fig. 
\ref{totgrads.fig} are often rather noisy, we can clearly see that in a
significant number of cases ($\approx 20$\%) the {\it non-zero} colour
gradients on the ``dust-free'' side of the galactic planes show
increasingly red colours with distance from the plane for profiles that
were extracted at progressively greater galactocentric distances.  This
may, in fact, be caused by the presence of young stars at small {\it z}
heights at these galactocentric distances, where active star formation
is ongoing.  In this case, we observe increasingly large vertical age
gradients at progessively greater galactocentric distances, while the
metallicity gradients might remain approximately similar.  Only in a few
cases we observe a general blueing with {\it z} height, for all of the
colour profiles of a given galaxy.  We will now investigate whether
these gradients correlate with global galaxy parameters. 

\subsection{Correlations with fundamental galaxy parameters}

\begin{figure*} 
\vspace*{-0.5cm} 
\psfig{figure=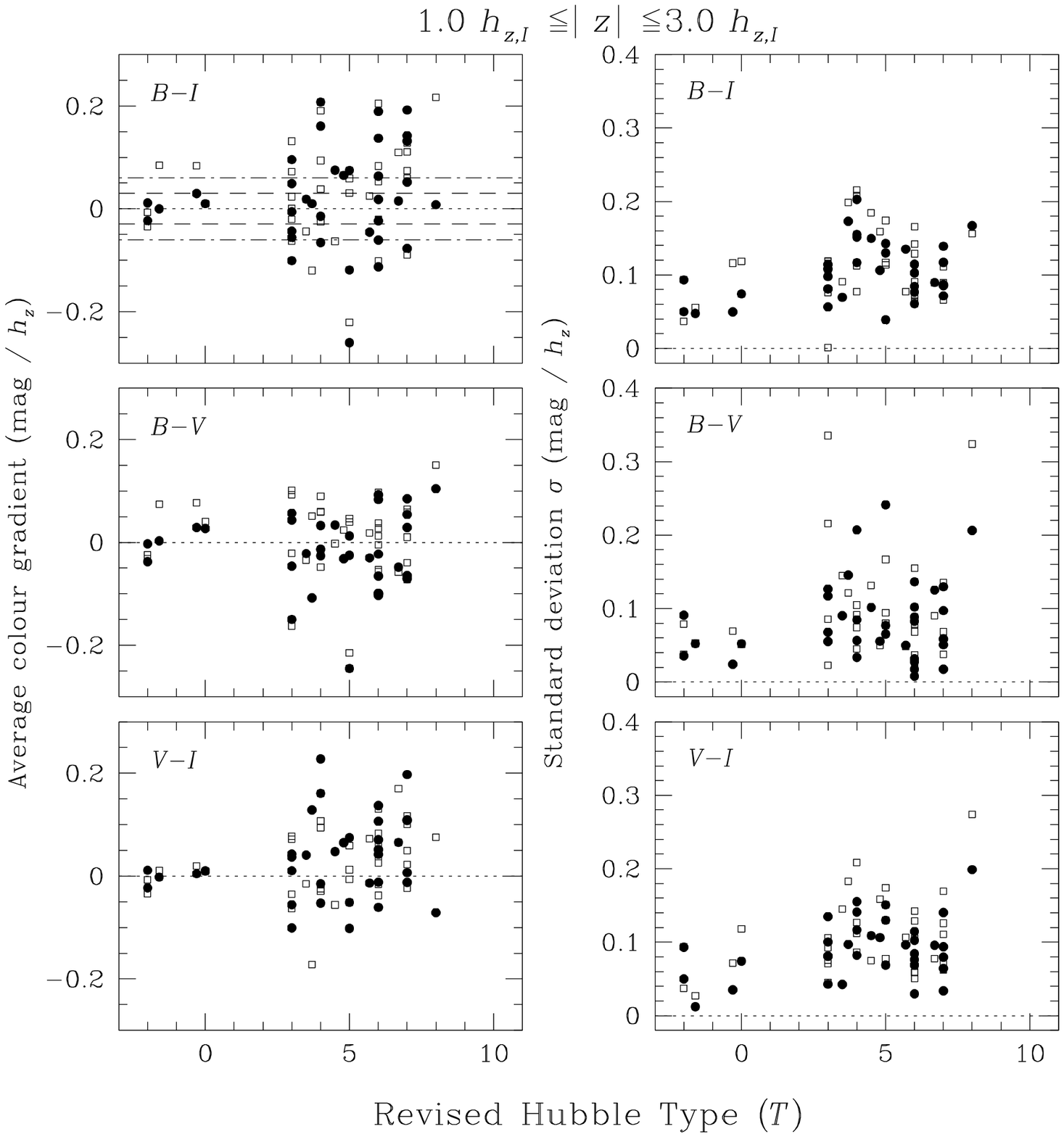,width=19cm}
\caption[]{\label{types.fig}Average vertical colour gradients and
typical scatter among gradients obtained from individual profiles in the
same galaxy as a function of galaxy type, {\it T}.  We have plotted the
results obtained from the dustiest side of the plane by open squares and
the side least affected by extinction by filled circles.  The fits were
done in the range $1.0 h_z \le |z| \le 3.0 h_z$, except for the galaxies
indicated in the caption of Fig.  \ref{totgrads.fig}.  The dashed lines
in the $(B-I)$ panel correspond to the vertical population gradient
expected for the Galaxy, governed by a constant SFR; the dash-dotted
lines are for an exponentially decaying SFR with an {\it e}-folding time
scale of 8 Gyr.}
\end{figure*}

Initial visual examination of the panels in Fig.  \ref{totgrads.fig}
hints at a possible dependence of the magnitude and scatter in the
vertical colour gradients on galaxy type.  Therefore, we determined the
average vertical colour gradient and its (1$\sigma$) standard deviation
for each of our galaxies, based on the results obtained in the previous
section.  The results of this exercise are presented in Fig. 
\ref{types.fig}, where we do indeed detect apparent correlations between
these parameters.  The correlations are in the sense that the range of
possible vertical colour gradients and their corresponding scatter is
larger for the spiral galaxy types ($T \ge 3$, or Sb) than for the
earliest, lenticular types ($T \le 0$, or S0). 

In de Grijs et al.  (1997) and de Grijs (1998) we found that, even at
{\it z} distances between 1.5 and $3.5 h_z$, luminosity profiles
parallel to the galaxies' minor axes are still non-negligibly affected
by dust extinction.  The distribution of vertical colour gradients in
Fig.  \ref{types.fig} is reminiscent of the distribution of the ratios
of {\it I} to {\it K}-band radial scale lengths in de Grijs et al. 
(1997) and de Grijs (1998) that indicated residual extinction effects in
the later-type sample galaxies. 

However, there does not appear to be any correlation between the average
colour gradients and their observational dispersions, for none of the
colours studied in this paper, and for neither of the sides of the
galactic planes.  This suggests that, for a given galaxy, residual dust
effects are not solely responsible for the observed scatter among the
magnitudes of the vertical colour profiles. 

We also examined possible correlations between the observed vertical
colour gradients and other fundamental galaxy parameters, like their
radial and vertical scale parameters, rotational velocities, and
absolute ({\it I}-band) magnitudes, but did not detect any clear trends
among these parameters.  

Finally, we investigated possible correlations between radial and
vertical colour gradients.  We used the ratios of the {\it B} and {\it
I}-band scale lengths of our sample galaxies as indicators of their
radial colour gradients (see de Grijs 1998 and references therein). 
However, other than the observation that galaxies without measurable
radial gradient generally do not exhibit any vertical $(B-I)$ colour
gradient either, we do not find any significant correlation between the
radial and vertical colour gradients, in any of the three colours
studied in this paper. This result is consistent with our suggestion
that the vertical colour gradients we measure are likely intrinsic,
while the radial colour gradients are mainly reflecting residual
extinction effects (de Grijs et al. 1997, de Grijs 1998).

\section{Stellar population gradients from broad-band colours}

Broad-band colours are relatively easy to obtain and are therefore the
most widely used colour diagnostics to date. Moreover, they immediately
reveal the approximate nature of a galaxy, which is to first order
determined by the dominant stellar population and dust content. 

The interpretation of the vertical colour gradients obtained in the
previous section depends predominantly on the fundamental question
whether one would expect any measurable vertical stellar population
gradient over the range used in our analysis, and if so, what its
magnitude would be.  In the following, we will base the answer to this
question on the observational evidence in our own Galaxy, due to
observational constraints (e.g., spatial resolution).

The dust-free colours of a composite stellar system (i.e., a galaxy) are
a function of its age and star formation history.  Thus, in order to
address this question, we first need to obtain the dependence of the
system's mean age on the height above the Galactic plane.  Although this
seems to be a rather fundamental question, surprisingly little work has
been done in this field.  

J{\o}nch-S{\o}rensen (1995) observed, for his sample of F and early
G-type stars, that all stars younger than 4 Gyr were found at $z <
500$pc, whereas the {\it minimum} age increases with increasing height
above the Galactic plane up to at least $z \simeq 2$ kpc, where the
minimum age is 4-5 Gyr.  He also shows that the logarithmic age
distribution for stars found at $1 < |z| < 2.5$ kpc can be represented
by a Gaussian distribution with a mean age of $\sim 8$ Gyr,
approximately constant with {\it z} height.  The width of the Gaussian
distribution is comparable to the approximate accuracy of his age
estimates for the faintest stars. 

Knude (1997) studied a statistically significant sample of young ($t <
1.7$ Gyr) sharply defined main sequence / subgiant A stars in the
direction of the North Galactic Pole, with a sample median age of 0.75
Gyr, and concluded that they show a very clear trend of mean age with
{\it z} height, increasing almost linearly to $\sim 0.75$ Gyr at $z
\simeq 150-200$ pc, after which the mean age distribution levels off and
remains approximately constant up to the completeness limit at 450 pc. 

Apart from these papers, no other reference has been made to the
Galactic mean age distribution as a function of height above the plane. 
Therefore, we will discuss two alternative and interdependent methods to
obtain this information:

\begin{enumerate}
\item by supplementing the vertical metallicity gradients discussed
above (Sect.  \ref{metgrads.sect}) with the Galactic age-metallicity
relation (AMR); and
\item by combining the age-velocity dispersion relation (AVR) with the
dependence of the velocity dispersion on {\it z} height.
\end{enumerate}

In either case, the observational results are controversial and show
large discrepancies, however.

\subsection{The age -- metallicity relation}

The well-known dependence of metal abundance on age of the stars and
open clusters in our Galaxy is one of the most important constraints on
chemical evolution theories, although, after almost two decades of
research in this field, a consensus on the exact correlation has not yet
been reached. 

\subsubsection{The solar neighbourhood AMR}

In a landmark paper, Twarog (1980) published the first detailed study of
the local (solar neighbourhood) AMR, in which he concluded that the
metallicity increased by a factor of about 4 between 13 and 4 Gyr ago,
and only slightly since then.  Although Carlberg et al.  (1985) claimed
to have detected a serious discrepancy with Twarog's (1980) AMR for the
younger stars, based on new metallicity and age calibrations, and
differences in their sample selection, modern determinations of the
solar neighbourhood AMR, based on high-accuracy observations of
homogeneous samples of representative Galactic disc tracers, agree
relatively well, within the large scatter about the relation (see Sect. 
\ref{scatter.sect}), with Twarog's (1980) original determination (e.g.,
Nissen, Edvardsson \& Gustafsson 1985; Knude, Schnedler Nielsen \&
Winther 1987; Lee, Ann \& Sung 1989; Meusinger, Reimann \& Stecklum
1991; Edvardsson et al.  1993; Ng \& Bertelli 1998; see Meusinger et al. 
1991 and Carraro, Ng \& Portinari 1998 for reviews), although Edvardsson
et al.'s (1993) AMR gives slightly lower abundances (by $\sim 0.1$ dex
in [Fe/H]) for stars in the age range from about 7 to 11 Gyr (see also
Rocha-Pinto \& Maciel 1997). 

Estimates of the slope for the solar neighbourhood AMR range from none
(Freeman 1991, for stars with ages between 3 and 10 Gyr), 0.02 dex
Gyr$^{-1}$ (Carlberg et al.  1985, $t_{\rm stars} \le 15$ Gyr; Lee et
al.  1989, $t_{\rm stars} \le 10$ Gyr), and 0.04 dex Gyr$^{-1}$
(Meusinger et al.  1991, $t_{\rm stars} \le 15$ Gyr;
J{\o}nch-S{\o}rensen 1995, $2 \le t_{\rm stars} \le 12$ Gyr) to 0.07 dex
Gyr$^{-1}$ (Edvardsson et al.  1993, $t_{\rm stars} \le 15$ Gyr; Carraro
et al.  1998, $t_{\rm stars} > 10$ Gyr; Ng \& Bertelli 1998, $t_{\rm
stars} > 10$ Gyr). The most recent slope determinations, which are based
on {\sl Hipparcos} data, indicate that there is no apparent slope in the
solar neighbourhood AMR for stars younger than 10 Gyr (Carraro et al.
1998; Ng \& Bertelli 1998); the slope determined for the older stars
agrees very well with Twarog's (1980) original determination.

\subsubsection{The scatter about the AMR}
\label{scatter.sect}

The most striking feature of these AMR determinations is the enormous
intrinsic scatter around the average relation, which is generally
obtained by binning in age intervals.  This scatter, for a given age
comparable to the overall increase in metallicity over the lifetime of
the Galaxy, makes the correlation for the younger stars rather weak. 
Although it has been found to be relatively easy to explain the average
trend by theoretical models involving infall of gas, either from the
intergalactic medium or from the Galactic thick disc (Sommer-Larsen \&
Yoshii 1990), at a rate of about half the star formation rate (SFR,
e.g., Twarog 1980; Carlberg et al.  1985; Strobel 1991; see Meusinger et
al.  1991 for a review; Pilyugin \& Edmunds 1996a,b), possibly combined
with radial inflow across the disc (Carlberg et al.  1985, but see Lacey
\& Fall 1985 and Meusinger et al.  1991), causing metal enrichment to
occur in the disc, it has proven no easy task to explain the observed
scatter about the relationship in terms of realistic physical processes. 

Several mechanisms have been suggested to be responsible for the
observed spread in metal abundance at any given age (see Carraro et al.
1998 for a review):

\begin{enumerate}

\item Diffusion of stellar orbits into the solar neighbourhood by
scattering off molecular clouds, transient spiral waves, or infalling
gas clouds, causing the mixing of stars originating from different
Galactocentric distances, where the stellar population might have
evolved at a different rate (e.g., Grenon 1987; Edvardsson et al.  1993;
Fran\c{c}ois \& Matteucci 1993; J{\o}nch-S{\o}rensen 1995).  However, even
after correcting the observations for the expected effect associated
with this mechanism as well as for any spatial abundance gradient, the
remaining scatter remains sufficiently large to require additional
mechanisms (Edvardsson et al. 1993; see also Carraro et al. 1998). In
addition, Pilyugin \& Edmunds (1996b) point out that this effect would
require a larger scatter for older stars compared to the younger
population, which is not supported by the observations.

\item Self-enrichment of the gas due to non-instantaneous mixing,
causing local inhomogeneities to arise in the metallicity distribution
(Malinie et al.  1993).  This could be triggered by the infall of gas,
or by sequential bursts of star formation (e.g., Edvardsson et al. 
1993; Pilyugin \& Edmunds 1996b; van den Hoek \& de Jong 1997; Ng \&
Bertelli 1998).  Carraro et al.  (1998) note that although both infall
of gas and sequential star formation are observed to occur in the solar
neighbourhood, the combination of both mechanisms is required to explain
the observed scatter.  Finally, Pilyugin \& Edmunds (1996b) argue that
an irregular rate of accretion of unenriched gas will more effectively
create the observed scatter. 

\item The admixture of stars associated with different Galactic
structures, e.g., the thin disc, thick disc and halo components, each
with its own local AMR (Strobel 1991; J{\o}nch-S{\o}rensen 1995). 
However, Pardi, Ferrini \& Matteucci (1994) and Carraro et al.  (1998)
argue that this mechanism is not attractive, because of the low density
of the thick disc and halo components in the solar neighbourhood. 

\end{enumerate}

\subsubsection{Beyond the solar neighbourhood}

An attempt to study the AMR outside the solar neighbourhood has been
made by J{\o}nch-S{\o}rensen (1995), who studied an {\it in situ} sample
of F and early G-type stars in six selected directions within the
Galactic disc, and at {\it z}-heights typically below 2 kpc, thereby
including both thin and thick disc stars.  He found a remarkably similar
{\it thin-disc} AMR ($z < 0.7$ kpc) to the local AMR discussed above,
with a slope of $\sim 0.04$ dex Gyr$^{-1}$ for stars in the age range
from 2 to 12 Gyr. 

Alternatively, the AMR for Galactic open clusters has been studied by a
number of authors (e.g., Strobel 1991; Friel \& Janes 1993; Carraro \&
Chiosi 1994; Carraro et al.  1998).  Open clusters are supposed to be
good tracers of the Galactic AMR, because their ages, abundances, and
positions can be estimated to high accuracy, and they represent the old
thin disc population (e.g., Strobel 1991; Friel \& Janes 1993). 

Although selection effects related to, among others, the cluster
survival mechanisms, their position in the Galactic disc and the
possible existence of a spatial metallicity gradient play a strong role
in the determination of the open cluster AMR (e.g., Friel \& Janes 1993;
Carraro \& Chiosi 1994; Carraro et al.  1998), the AMRs for open
clusters and stars are essentially the same, including the large scatter
about the relationship. 

A similar relationship as for the solar neighbourhood stars has been
found for the open clusters in the Galactic anticenter direction (Friel
\& Janes 1993) as well as in the LMC (e.g., Olzewski et al.  1991;
Strobel 1991; Friel \& Janes 1993), although chemical enrichment has
taken place at a slower rate in the LMC than in the Galaxy. 

\subsection{The age -- vertical velocity dispersion relation}

Although it is well-known that the velocity dispersions of disc stars
vary with age, the lack of accurate kinematical observations has
prevented the detailed study of the correlation between the mean stellar
age and their vertical velocity dispersions (referred to AVR{\it z}
henceforth).  In addition, a number of discrepant AVR{\it z}'s has
appeared in the literature (e.g., Mayor 1974; Wielen 1977; Carlberg et
al.  1985; Str\"omgren 1987; Freeman 1991; Meusinger et al.  1991;
G\'omez et al.  1997; see Haywood, Robin \& Cr\'ez\'e 1997a,b for a
comparison), adding confusion to the already obscure picture. 

The AVR of stars in the solar neighbourhood traces the kinematical and
dynamical evolution of the local Galactic disc and is an indicator of
the heating mechanism by which the stellar random velocities are
increased after their birth.  This diffusion of the stellar orbits in
the Galactic disc is caused by irregularities in the Galactic
gravitational field (Wielen 1977; Wielen \& Fuchs 1983), like the
interaction with giant molecular clouds (e.g., Spitzer \& Schwarzschild
1951, 1953; Lacey 1984) or recurrent transient spiral waves (e.g.,
Barbanis \& Woltjer 1967; Carlberg \& Sellwood 1985).  Both mechanisms
predict that the stellar velocity dispersions are a continuous function
of age, such that for young stars $\hbox{d}\sigma_W^2 / \hbox{d}t
\approx$ constant (e.g., Carlberg et al.  1985), where $\sigma_W^2$
represents the vertical velocity dispersion in the Galactic disc.  For
older stars, $\hbox{d}\sigma_W^2 / \hbox{d}t \approx \sigma_W^{-2}$. 

Wielen's (1977) often referenced AVR showed a smooth increase of the
stellar velocity with age, although for solar abundance stars the
vertical velocity dispersion seems to become approximately constant in
the age range from about 3-6 to 13 Gyr (e.g., Carlberg et al.  1985;
Str\"omgren 1987; Freeman 1991; G\'omez et al.  1997; Haywood et al. 
1997a,b). 

With the recent acquisition of highly accurate kinematical data and more
accurate age dating techniques, it has become possible to study the
features of the AVR{\it z} in more detail.  Freeman (1991) argues that
his AVR{\it z}, which is based on Edvardsson et al.'s (1993) new
kinematically unbiased sample of nearby F and early G-type stars for
which ages are known with high precision, appears to show three age
zones in terms of the stellar vertical velocity dispersions, $t < 3$
Gyr, $3 \le t \le 10$ Gyr, and $t > 10$ Gyr.  A similar conclusion is
reached by G\'omez et al.  (1997), based on new observations of 2812
stars in the {\sl Hipparcos} Catalogue (ESA 1997).  G\'omez et al. 
(1997) subdivide the youngest age bin into two subbins, i.e., stars with
$t < 0.8 - 1$ Gyr, corresponding to incomplete mixing, and stars between
about 1 and $3-4$ Gyr. 

The second age bin, which is characterized by the saturation of the
vertical velocity dispersion at about 15-17 km s$^{-1}$ (G\'omez et al. 
1997) to $19 \pm 2$ km s$^{-1}$ (Freeman 1991), does not show any
correlation of vertical velocity dispersion with age, and represents the
old-disc population (Freeman 1991).  Haywood et al.  (1997a) split up
the old-disc component into three further subdivisions (3-5, 5-7 and
7-10 Gyr) and caution that on neither the shape nor the level at which
the AVR{\it z} saturates consensus has yet been reached.  However, it
appears that the disc heating saturates after 3-6 Gyr, because there
seems to be no dynamical disc evolution between 3-6 and 10 Gyr. 

The kinematics and metallicities of the stars in the highest-age bin
indicate that they belong to the thick disc (Freeman 1991).

\subsection{The dependence of mean age on height above the Galactic plane}

From the discussion of the AMR, we adopt for the thin-disc ages
associated with the stars in the vertical range of interest (between
about 4 and 10 Gyr; Freeman 1991) the mean slope of 0.04 ($\pm 0.03$)
dex Gyr$^{-1}$, which also seems to apply to the first two scale heights
of the old-disc population towards the North Galactic Pole
(J{\o}nch-S{\o}rensen 1995), and is consistent with the most recent and
highest-precision results. 

As we discussed in Sect.  \ref{metgrads.sect}, most modern estimates of
the vertical metallicity gradient in the Galactic disc are in the range
$-0.6 \le \hbox{\rm d[Fe/H]/d{\it z}} \le -0.3$ dex kpc$^{-1}$. 
However, J{\o}nch-S{\o}rensen's (1995) best results for $z < 700$ pc and
solved simultaneously for both a radial and a vertical gradient,
indicate that the {\it in situ} vertical gradient is of order $-0.2 \pm
0.3$ dex kpc$^{-1}$, of which the error estimate partially overlaps with
the range quoted before.  Since J{\o}nch-S{\o}rensen's (1995) analysis
was done in a similar vertical range as that where we measured our
colour gradients, and because he also properly accounted for the
presence of possible radial colour gradients, we will adopt his result
in the following analysis. 

If we combine both dependences,
\begin{equation}
\hbox{[Fe/H]} = + 0.04 (\pm 0.03) \times \Big( {t \over \hbox{Gyr}} \Big)
+ C ,
\end{equation}
for $4 \le t \le 10$ Gyr, and
\begin{equation}
\hbox{[Fe/H]} = -0.2 (\pm 0.3) \times \Big( {z \over \hbox{kpc}} \Big) +
C' ,
\end{equation}
we derive
\begin{equation}
\label{agez.eq}
\Big( {t \over \hbox{Gyr}} \Big) = 5.0 (\pm 8.4) \times \Big( {z \over
\hbox{kpc}} \Big) + C^* ,
\end{equation}
for $4 \le t \le 10$ Gyr.

To compare this dependence for the Galaxy to our results, obtained at
{\it z} heights between $1.0 h_z \le |z| \le 3.0 h_z$, we need to know
the Galactic thin disc scale height.  From the overview by Sackett
(1997), it appears that the most recent estimates of the Galactic thin
disc scale height in the solar neighbourhood converge at $0.26 \pm 0.05$
kpc (Ojha et al.  1996; see also Kent, Dame \& Fazio 1991).  Thus, we
should compare our results from external edge-on galaxies with the
region $0.26 \le |z| \le 0.78$ kpc in the Galaxy; over this {\it z}
range, the mean age of the dominant population increases only by about
2.6 Gyr (Eq.  (\ref{agez.eq})); as a reminder, the mean age of the stars
in the Galactic old thin disc for which J{\o}nch-S{\o}rensen (1995)
obtained these relations is about 8 Gyr. 

If we combine the most recent results on the AVR with the observation
that the stellar velocity dispersion in the radial direction of A stars
in the direction of the North Galactic Pole increases up to about 300 pc
and might flatten out at higher {\it z} heights (although the data is
rather noisy; Knude 1997), the results from the AVR analysis are
consistent with the AMR data. 

Now, we can combine this information with stellar population synthesis
models of composite galactic systems, like the ones provided by Bruzual
\& Charlot (1993; henceforth BC93).  The colours of a galactic stellar
population at ages greater than about 0.2 Gyr depend sensitively on the
star formation history assumed (BC93).  Although the precise star
formation histories of spiral galaxies are still heavily debated,
estimates of the global star formation history over the lifetime of the
Galaxy are generally consistent with a constant SFR (e.g., Twarog 1980;
Meusinger et al.  1991). 

\begin{figure*} 
\vspace*{-0.5cm} 
\psfig{figure=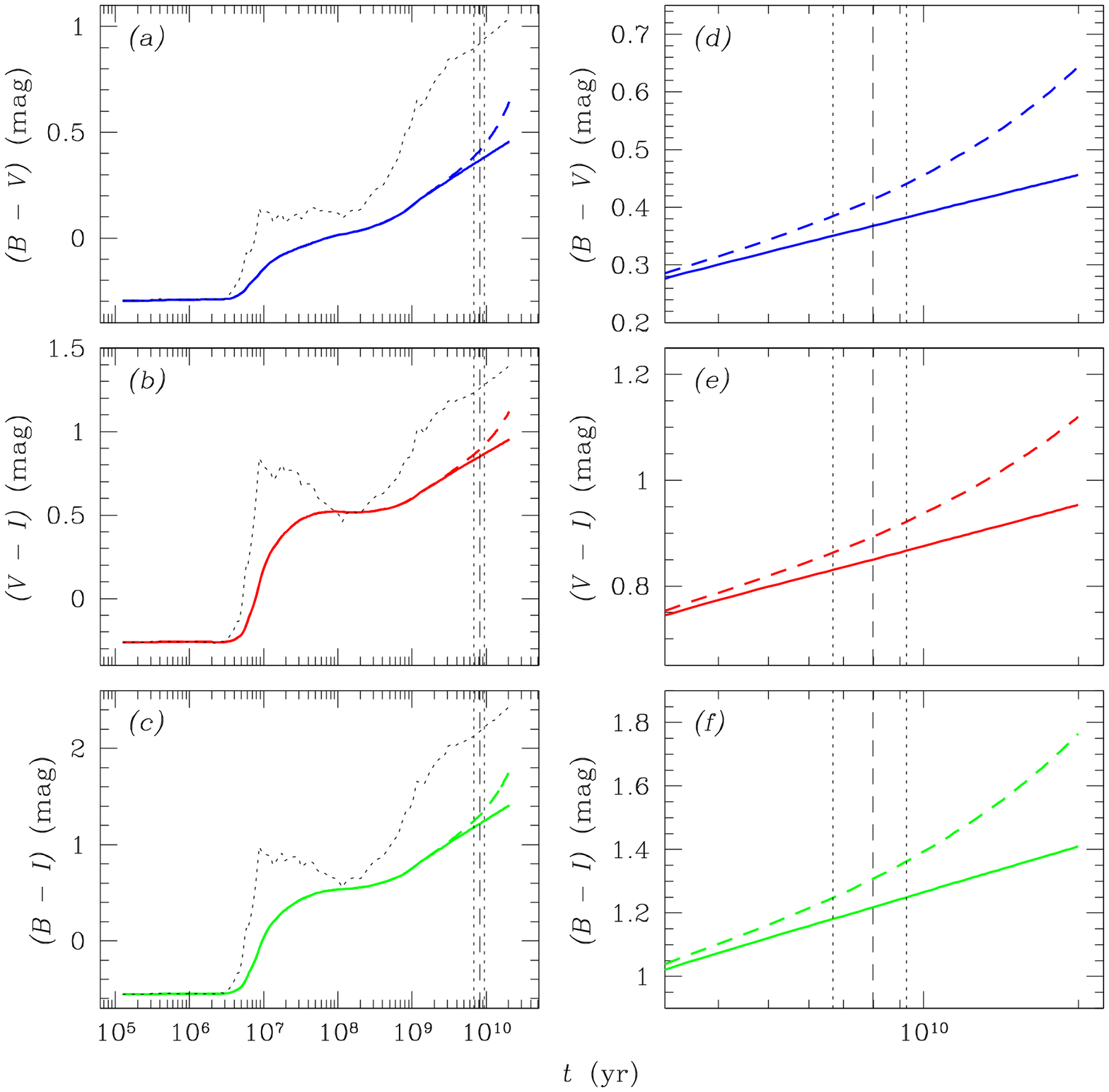,width=19cm}
\caption[]{\label{BCcolors.fig}Time dependence of our broad-band colours
for a composite stellar population (BC93).  Solid lines: constant SFR;
dashed lines: exponentially decaying SFR with an {\it e}-folding time
scale of 8 Gyr; thin dotted lines: single burst colour evolution.  {\it
(a)--(c)} Evolution over the system's entire lifetime; {\it (d)--(f)}
close-ups of the age range currently dominated by the Galactic old thin
disc.  The mean age of the stars in this Galactic substructure, 8 Gyr
(J{\o}nch-S{\o}rensen 1995), has been indicated by the vertical dashed
lines; the vertical dotted lines delineate a 2.6 Gyr age range about
this mean age (see text).}
\end{figure*}

Under these assumptions, the stellar population differences expected to
manifest themselves in our broad-band colours, for an age range of 2.6
Gyr centered on the mean old-disc age of 8 Gyr (J{\o}nch-S{\o}rensen
1995), amount to $\Delta (B-V) \approx 0.03, \Delta (V-I) \approx 0.04$,
and $\Delta (B-I) \approx 0.07$ mag.  The corresponding population
gradients are $\simeq 0.02, 0.02$, and 0.03 mag $h_{z,I}^{-1}$ in
$(B-V), (V-I),$ and $(B-I)$, respectively.  In Fig.  \ref{BCcolors.fig}
we show the evolution of our broad-band colours for a composite stellar
population, based on the convolution integrals over time of the spectral
energy distributions for a single stellar population of solar
metallicity (BC93), for both the generally assumed constant SFR (solid
lines) and an exponentially decaying SFR with an {\it e}-folding time
scale of 8 Gyr, according to Pilyugin \& Edmunds' (1996a) best-fitting
model to describe the Galactic AMR (dashed lines).  To compute the
colours in the time-dependent SFR case, we have assumed ``closed box''
evolution of the composite stellar population, i.e., without the infall
of extragalactic gas, but including gas recycling (and metal enrichment)
with an efficiency of 46\% (Haywood et al.  1997b; see also Kennicutt,
Tamblyn \& Congdon 1994) until the present time, and assuming a Salpeter
(1955) initial mass function for stellar masses ranging from 0.1 to 125
$M_\odot$.  If we assume that the exponentially decaying SFR advocated
by Pilyugin \& Edmunds (1996a) is a more representative description of
the Galactic evolution, these broad-band vertical colour gradients due
to population gradients increase to $\Delta (B-V) \simeq 0.03$ mag
$h_{z,I}^{-1}, \Delta (V-I) \simeq 0.03$ mag $h_{z,I}^{-1}$, and $\Delta
(B-I) \simeq 0.06$ mag $h_{z,I}^{-1}$.  Finally, for comparison, we have
also included the canonical single-burst colour evolution (thin dotted
lines in Figs.  \ref{BCcolors.fig}a--c; BC93).  The colour gradients for
these models are similar to those obtained from the models of composite
stellar systems with a constant SFR, although these single-burst colours
become significantly redder over their lifetimes. 

For simplicity, we will assume that these population gradients, and the
star formation histories assumed to obtain them, are applicable to all
our sample galaxies and are representative of their stellar populations,
irrespective of galaxy type.  If we examine Fig.  \ref{types.fig} in
detail, and in particular the panel corresponding to the average
vertical colour gradients in the $(B-I)$ images, where any population
gradients would show up most prominently, it turns out that about 17 and
26, or about 46 and 70\% of our 37 sample galaxies exhibit average
vertical colour gradients in excess of the population gradients expected
from, respectively, an exponentially decaying SFR as discussed above, or
a constant SFR. 

Even though one should be cautious in interpreting these percentages,
due to uncertainties in the model assumptions, it is obvious that for a
significant fraction of our sample galaxies another mechanism is
required to explain the magnitude of the observed gradients, like, e.g.,
residual dust extinction at the {\it z} heights studied in this paper
(Sect.  \ref{vertcols.sect}) or vertical metallicity gradients. 

\section{Vertical metallicity gradients}

The colour and metallicity distributions in a galactic disc are the
result of the composite star formation and chemical enrichment history
of the galaxy over its lifetime, and not a representation of the
composition of a single stellar population in the galactic disc at any
given time.  As J{\o}nch-S{\o}rensen (1995) argues, an evolution of the
metallicity gradient with time may be expected because of the gas flows
in a galaxy and also because of the dependence of chemical enrichment
on, e.g., the gas fraction of the total mass and the star formation
efficiency.  Therefore, since the observed gradients are also affected
by the disc dynamical evolution (i.e., orbit diffusion), the original
gradients may be very hard to trace for stellar populations that are
older than a few Gyr. 

It is expected that, over the lifetimes of galaxies, galactic discs
accrete a substantial number of lumps of stars, gas and dark matter, or
satellite galaxies, through infall (cf.  the Searle \& Zinn [1978]
Galaxy formation hypothesis).  The effects of such minor mergers are not
clear, however.  Norris \& Ryan (1989) and Norris \& Green (1989) argue
that such random merger events cannot explain the continuous metallicity
behaviour as a function of height above the plane.  However, Quinn,
Hernquist \& Fullagar (1993) argue that the correlations of abundance
with scale height and vertical velocity are not destroyed by mergers;
they show, based on numerical simulations, that an initial abundance
gradient with {\it z} height can be accentuated by a merger, as is also
seen in mergers of spherical systems and cosmological models of halo
formation (see Quinn et al.  1993 for a review).  On the other hand,
Gilmore et al.  (1995) argue that the dissipationless relaxation of the
thin-disc constituents after a minor merger will weaken any existing
metallicity gradient, but not destroy it completely (e.g., White 1980;
Hernquist \& Quinn 1993). 

The analysis of the present sample may help distinguish between these
scenarios.  It is generally assumed that lenticular (S0) galaxies have
undergone a greater of number significant interactions or accretion
events than later-type spiral galaxies (see, e.g., Schweizer [1996] for
a review).  Combined with the observational evidence from Fig. 
\ref{types.fig}, this suggests that initial vertical metallicity
gradients, if any, have likely not been accentuated by these accretion
or merging events.  On the other hand, they may have weakened any
existing vertical gradient in metal abundance, although we cannot rule
out the possibility that they have left the existing correlations
unchanged.  Of course, this analysis depends critically on the
assumptions made regarding the uniformity of the initial metallicity
gradients as a function of galaxy type.  To date, no models have been
constructed that include the treatment of metal abundance during galaxy
formation, however. 

\section{Summary and conclusions}

We have analyzed optical {\it B, V}, and {\it I}-band
observations and their corresponding colour maps of a statistically
complete sample of edge-on disc-dominated spiral and lenticular
galaxies.  The main aim of this study was to examine the intrinsic
vertical colour gradients in the discs of our sample galaxies, and to
constrain the effects of population gradients, residual dust extinction
and gradients in the galaxies' metal abundance, based on the observed
changes of broad-band colour with increasing height above the galactic
planes. 

For our current study, we needed to isolate those vertical ranges of our
edge-on galactic discs that were as little as possible affected by
reddening due to extinction, poor S/N ratios, foreground stars, or
out-of-plane spiral arms.  We concluded that, for the majority of our
sample galaxies, the colours and colour gradients in the range $1.0 h_z
\le |z| \le 3.0 h_z$ most likely reflect the intrinsic galactic
properties, since any small-scale variations in the vertical colour
profiles due to external effects are smoothed out over this relatively
large range. 

From the detailed examination of the vertical colour profiles in our
sample galaxies (Figs.  \ref{totgrads.fig} and \ref{colprofs.fig}), it
appears that the intrinsic vertical colour gradients are either
non-existent, or small and relatively constant as a function of position
along the galaxies' major axes.  On average, the earlier-type galaxies
exhibit smaller vertical $(B-I)$ gradients (and scatter among the
average gradients) than the later types; our results are consistent with
the absence of any vertical colour gradient in the discs of our
early-type sample galaxies. 

In most galaxies small-scale variations in the magnitude and even the
direction of the vertical gradient are observed, in the sense that at
larger galactocentric distances they generally display redder colours
with increasing {\it z} height, whereas the opposite is often observed
in and near the galactic centres. 

About half to two-thirds of our sample galaxies exhibit average vertical
colour gradients in excess of the population gradients expected from
observational data of the Galaxy, based on the SFR assumed.  Although
one should be cautious in interpreting these results directly, due to
the numerous model assumptions involved, it is obvious that for a
significant fraction of our sample galaxies another mechanism is
required to explain the magnitude of the observed gradients, like, e.g.,
residual dust extinction or vertical metallicity gradients. 

The non-zero colour gradients in a significant fraction of our sample
galaxies are likely (at least) partially due to residual dust extinction
at these {\it z} heights, as is also evidenced from the sometimes
significant differences between the vertical colour gradients measured
on either side of the galactic planes.  In addition, in the cases where
we observe a blueing of the galactic disc with height above the plane,
we cannot always rule out the possibility that this is due to extinction
effects.

Finally, several scenarios have been proposed to describe the evolution
of a vertical metallicity gradient over the lifetime of a galaxy, under
the influence of infall or minor merger events.  It is generally assumed
that lenticular galaxies have undergone a greater of number significant
interactions or accretion events than later-type spiral galaxies. 
Combined with the observational evidence from Fig.  \ref{types.fig},
this suggests that initial vertical metallicity gradients, if any, have
likely not been accentuated by these accretion or merging events, as
suggested by Quinn et al.  (1993).  On the other hand, they may have
weakened any existing vertical gradient in metal abundance (Gilmore et
al.  1995), although we cannot rule out the possibility that they have
left the existing correlations unchanged. 

\section*{Acknowledgments} During most of this work RdeG was supported
by NASA grants NAG 5-3428 and NAG 5-6403.  RdeG also wishes to thank the
Department of Physics of the University of Durham for their hospitality
on two visits.  This research has made use of NASA's Astrophysics Data
System Abstract Service.


\begin{thebibliography}{}

\bibitem[]{} Aoki T.E., Hiromoto N., Takami H., Okamura S., 1991, PASJ,
43, 755
\bibitem[]{} Barteldrees A., Dettmar R.-J., 1994, A\&AS, 103, 475
\bibitem[]{} Bahcall J.N., Soneira R.M., 1980, ApJS, 44, 73
\bibitem[]{} Barbanis B., Woltjer L., 1967, ApJ, 150, 461 
\bibitem[]{} Bruzual A., G., Charlot S., 1993, ApJ, 405, 538
\bibitem[]{} Buser R., Rong J.X., 1995, Baltic Astron., 4, 1
\bibitem[]{} Buser R., Rong J.X., Karaali S., 1998, A\&A, 331, 934
\bibitem[]{} Carlberg R.G., Dawson P.C., Hsu T., vandenBerg D.A., 1985,
ApJ, 294, 674
\bibitem[]{} Carlberg R.G., Sellwood J.A., 1985, ApJ, 292, 79
\bibitem[]{} Carraro G., Chiosi C., 1994, A\&A, 287, 761
\bibitem[]{} Carraro G., Ng Y.K., Portinari L., 1998, MNRAS, 296, 1045
\bibitem[]{} Chen B., 1997, AJ, 113, 311
\bibitem[]{} Davies J.I., 1990, MNRAS, 244, 8
\bibitem[]{} de Grijs R., 1997, PhD Thesis, Univ.  Groningen, the
Netherlands
\bibitem[]{} de Grijs R., 1998, MNRAS, 299, 595
\bibitem[]{} de Grijs R., Peletier R.F., 1999, MNRAS, 310, 157
\bibitem[]{} de Grijs R., Peletier R.F., van der Kruit P.C., 1997, A\&A,
327, 966
\bibitem[]{} de Jong R.S., 1996, A\&A, 313, 377
\bibitem[]{} de Jong R.S., van der Kruit P.C., 1994, A\&AS, 106, 451
\bibitem[]{} de Vaucouleurs G., de Vaucouleurs A., Corwin H.G., Jr.,
Buta R.J., Paturel G., Fouqu\'e P., 1991, Springer-Verlag, New York (RC3)
\bibitem[]{} Edvardsson E., Andersen J., Gustafsson B., Lambert D.L.,
Nissen P.E., Tomkin J., 1993, A\&A, 275, 101
\bibitem[]{} ESA, 1997, The {\sl Hipparcos} and {\sl Tycho} Catalogue,
ESA SP-1200
\bibitem[]{} Fisher D., Franx M., Illingworth G., 1996, ApJ, 459, 110
\bibitem[]{} Fran\c{c}ois P., Matteucci F., 1993, A\&A, 280, 136
\bibitem[]{} Freeman K.C., 1991, in Sundelius B., ed., Dynamics of Disc
Galaxies.  G\"oteborg, p.  15
\bibitem[]{} Friel E.D., Janes K.A., 1993, A\&A, 267, 75
\bibitem[]{} Gilmore G., Wyse R.F.G., Jones J.B., 1995, AJ, 109, 1095
\bibitem[]{} G\'omez A.E., Grenier S., Udry S., Haywood M., Meillon L.,
Sabas V., Sellier A., Morin D., 1997, in ESA SP-402, Hipparcos -- Venice
1997.  Venice, p.  621
\bibitem[]{} Grenon M., 1977, in M\"uller E.A., ed., Highlights of
Astronomy Vol.  4 Part {\sc ii}. IAU, p.  55
\bibitem[]{} Grenon M., 1987, J. Astrophys. Astr., 8, 123
\bibitem[]{} Hamabe M., Kodaira K., Okamura S., Takase B., 1979, PASJ,
31, 431
\bibitem[]{} Hamabe M., Kodaira K., Okamura S., Takase B., 1980, PASJ,
32, 197
\bibitem[]{} Hartkopf W.I., Yoss K.M., 1982, AJ, 87, 1679
\bibitem[]{} Haywood M., Robin A.C., Cr\'ez\'e M., 1997a, A\&A, 320, 428
\bibitem[]{} Haywood M., Robin A.C., Cr\'ez\'e M., 1997b, A\&A, 320, 440
\bibitem[]{} Hegyi D.J., Gerber G., 1979, in Evans D.S., ed.,
Photometry, Kinematics, and Dynamics of Galaxies.  Univ.  of Texas
Astron.  Dept., Austin, p.  119
\bibitem[]{} Hernquist L., Quinn P.J., 1993, in Majewski S.R., ed., ASP
Conf.  Ser.  Vol.  49, Galaxy Evolution -- The Milky Way Perspective,
Astron.  Soc.  Pac., San Francisco, p.  187
\bibitem[]{} Jansen R.A., Knapen J.H., Beckman J.E., Peletier R.F., Hes
R., 1994, MNRAS, 270, 373
\bibitem[]{} Jensen E.B., Thuan T.X., 1982, ApJS, 50, 421
\bibitem[]{} J{\o}nch-S{\o}rensen H., Knude J., 1994, A\&A, 288, 139
\bibitem[]{} J{\o}nch-S{\o}rensen H., 1995, A\&A, 298, 799
\bibitem[]{} Kennicutt Jr.  R.C., Tamblyn P., Congdon C.W., 1994, ApJ,
435, 22
\bibitem[]{} Kent S.M., Dame T.M., Fazio G., 1991, ApJ, 378, 131
\bibitem[]{} Knude J., 1997, A\&A 327, 90
\bibitem[]{} Knude J., Schnedler Nielsen H., Winther M., 1987, A\&A, 179,
115
\bibitem[]{} Kuchinski L.E., Terndrup D.M., 1996, AJ, 111, 1073
\bibitem[]{} Lacey C.G., 1984, MNRAS, 208, 687
\bibitem[]{} Lacey C.G., Fall S.M., 1985, ApJ, 290, 154
\bibitem[]{} Lauberts A., Valentijn E.A., 1989, The Surface Photometry 
Catalogue of the ESO-Uppsala Galaxies. ESO, Garching bei M\"unchen (ESO-LV)
\bibitem[]{} Lee S.-W., Ann H.B., Sung H., 1989, JKAS, 22, 43
\bibitem[]{} Lequeux J., Fort B., Dantel-Fort M., Cuillandre J.-C.,
Mellier Y., 1996, A\&A, 312, L1
\bibitem[]{} Lequeux J., Combes F., Dantel-Fort M., Cuillandre J.-C.,
Fort B., Mellier Y., 1998, A\&A, 334, L9
\bibitem[]{} Majewski S.R., 1992, ApJS, 78, 87
\bibitem[]{} Malinie G., Hartmann D.H., Clayton D.D., Mathews G.J.,
1993, ApJ, 413, 633
\bibitem[]{} Mathewson D.S., Ford V.L., Buchhorn M., 1992, ApJS, 81, 413
\bibitem[]{} Mathewson D.S., Ford V.L., 1996, ApJS, 107, 97
\bibitem[]{} Mayor M., 1974, A\&A, 32, 321
\bibitem[]{} Meusinger H., Reimann H.-G., Stecklum B., 1991, A\&A, 245, 57
\bibitem[]{} Ng Y.K., Bertelli G., 1998, A\&A, 329, 943
\bibitem[]{} Nissen P.E., Edvardsson B., Gustafsson B., 1985, in
Danziger I.J., Matteucci F., Kj\"ar K., eds., Proc.  ESO Workshop on
Production and Distribution of CNO Elements.  ESO, Garching bei
M\"unchen, p.  131
\bibitem[]{} Norris J.E., Green E.M., 1989, ApJ, 337, 272
\bibitem[]{} Norris J.E., Ryan S.G., 1989, ApJ, 336, L17
\bibitem[]{} Ojha D.K., Bienaym\'e O., Robin A.C., Cre\'z\'e M., Mohan
V., 1996, A\&A, 311, 456
\bibitem[]{} Olzewski E.W., Schommer R.A., Suntzeff N.B., Harris H.C.,
1991, AJ, 101, 515
\bibitem[]{} Pardi M.C., Ferrini F., Matteucci F., 1994, ApJ, 444, 207
\bibitem[]{} Peletier R.F., Balcells M., 1996, AJ, 111, 2238
\bibitem[]{} Peletier R.F., Balcells M., 1997, NewA, 1, 349
\bibitem[]{} Pilyugin L.S., Edmunds M.G., 1996a, A\&A, 313, 783
\bibitem[]{} Pilyugin L.S., Edmunds M.G., 1996b, A\&A, 313, 792
\bibitem[]{} Quinn P.J., Hernquist L., Fullagar D.P., 1993, ApJ, 403, 74
\bibitem[]{} Reid N., Majewski S.R., 1993, ApJ, 409, 635
\bibitem[]{} Robin A.C., Haywood M., Cr\'ez\'e M., Ojha D.K., Bienaym\'e
O., 1996, A\&A, 305, 125
\bibitem[]{} Rocha-Pinto H.J., Maciel W.J., 1997, MNRAS, 289, 882
\bibitem[]{} Rudy R.J., Woodward C.E., Hodge T., Fairfield S.W., Harker
D.E., 1997, Nat., 387, 159
\bibitem[]{} Sackett P.D., 1997, ApJ, 483, 103
\bibitem[]{} Salpeter E.E., 1955, ApJ, 121, 161
\bibitem[]{} Sasaki T., 1987, PASJ, 39, 849
\bibitem[]{} Schmidt M., 1963, ApJ, 137, 758
\bibitem[]{} Schweizer F., 1996, in Kennicutt Jr.  R.C., Schweizer F.,
Barnes J.E., eds., Saas-Fee Advanced Course 26, Galaxies: Interactions
and Induced Star Formation.  Springer-Verlag, Heidelberg, p.  105
\bibitem[]{} Searle L., Zinn R., 1978, ApJ, 225, 357
\bibitem[]{} Sommer-Larsen J., Yoshii Y., 1990, MNRAS, 243, 468
\bibitem[]{} Spitzer L., Schwarzschild M., 1951, ApJ, 114, 385
\bibitem[]{} Spitzer L., Schwarzschild M., 1953, ApJ, 118, 106
\bibitem[]{} Strobel A., 1991, A\&A, 247, 35
\bibitem[]{} Str\"omgren B., 1987, in Gilmore G., Carswell R., eds.,
Proc.  NATO Adv.  Study Inst., The Galaxy.  Reidel, Dordrecht, p.  229
\bibitem[]{} Tinsley B., 1980, Fund. Cosmic Phys., 5, 287
\bibitem[]{} Trefzger C.F., Pel J.W., Gabi S., 1995, A\&A, 304, 381
\bibitem[]{} Twarog B.A., 1980, ApJ, 242, 242
\bibitem[]{} van den Berg S., 1962, AJ, 67, 1165
\bibitem[]{} van den Hoek L.B., de Jong T., 1997, A\&A, 318, 231
\bibitem[]{} van der Kruit P.C., A\&A, 192, 117
\bibitem[]{} van der Kruit P.C., Searle L., 1981, A\&A, 95, 116
\bibitem[]{} van der Kruit P.C., Searle L., 1982a, A\&A, 110, 61
\bibitem[]{} van der Kruit P.C., Searle L., 1982b, A\&A, 110, 79
\bibitem[]{} Wainscoat R.J., Freeman K.C., Hyland A.R., 1989, ApJ, 337,
163
\bibitem[]{} White S.D.M., 1980, MNRAS, 191, 1P
\bibitem[]{} Wielen R., 1977, A\&A, 60, 263
\bibitem[]{} Wielen R., Fuchs B., 1983, in Shuter W.L., ed., Kinematics,
Dynamics and Structure of the Milky Way.  Reidel, Dordrecht, p.  81
\bibitem[]{} Worthey G., 1994, ApJS, 95, 107
\bibitem[]{} Yoshii Y., Ishida K., Stobie R.S., 1987, AJ, 93, 323
\bibitem[]{} Yoss K.M., Neese C.L., Hartkopf W.I., 1987, AJ, 94, 1600

\end{thebibliography}
\end{document}